\documentclass[twocolumn,showpacs,preprintnumbers,amsmath,amssymb,floatfix,superscriptaddress,PRC]{revtex4}
\usepackage{graphicx}
\usepackage{dcolumn}
\usepackage{bm}

  \def\nuc#1#2{\relax\ifmmode{}^{#1}{\protect\text{#2}}\else${}^{#1}$#2\fi}
  \def\itnuc#1#2{\setbox\@tempboxa=\hbox{\scriptsize\it #1}
    \def\@tempa{{}^{\box\@tempboxa}\!\protect\text{\it #2}}\relax
    \ifmmode \@tempa \else $\@tempa$\fi}

\begin{document}
\preprint{APS/123-QED}

\title{Neutron activation of natural zinc samples at $kT=25$~keV}

\author{R.~Reifarth}
\affiliation{Goethe Universit{\"a}t Frankfurt, Frankfurt a.M, 60438, Germany}
\author{S.~Dababneh}
\affiliation{Department of Physics, Faculty of Science, Al-Balqa Applied University, P.O. Box 2587, Amman 11941, Jordan}
\author{M.~Heil}
\affiliation{GSI Helmholtzzentrum f\"ur Schwerionenforschung GmbH, Darmstadt, 64291, Germany}
\author{F.~K\"appeler}
\affiliation{Karlsruhe Institute of Technology, Campus Nord, Institut f\"ur Kernphysik, 76021 Karlsruhe, Germany}
\author{R.~Plag}
\affiliation{Goethe Universit{\"a}t Frankfurt, Frankfurt a.M, 60438, Germany}
\author{K.~Sonnabend}
\affiliation{Goethe Universit{\"a}t Frankfurt, Frankfurt a.M, 60438, Germany}
\author{E.~Uberseder}
\affiliation{University of Notre Dame, Department of Physics, Notre Dame, Indiana, USA}

\date{\today}
\begin{abstract}
The neutron capture cross sections of  $^{64}$Zn,  $^{68}$Zn, and  $^{70}$Zn 
have been measured with the activation technique in a quasi-stellar neutron spectrum
corresponding to a thermal energy of $kT = 25$ keV. By a series of repeated 
irradiations with different experimental conditions, an uncertainty of 3\% could be 
achieved for the $^{64}$Zn($n,\gamma$)$^{65}$Zn cross section and for the
partial cross section $^{68}$Zn($n,\gamma$)$^{69}$Zn$^{m}$ feeding the
isomeric state in $^{69}$Zn. For the partial cross sections $^{70}$Zn($n,\gamma$)$^{71}$Zn$^{m}$ 
and $^{70}$Zn($n,\gamma$)$^{71}$Zn$^{g}$, 
which had not been measured so far,  uncertainties of only 16\% and 6\% could
be reached because of limited counting statistics and decay intensities. 
Compared to previous measurements on $^{64,68}$Zn, the uncertainties 
could be significantly improved, while the $^{70}$Zn cross section was found 
two times smaller than existing model calculations. From these results Maxwellian 
average cross sections were determined between 5 and 100~keV. Additionally, the 
$\beta$-decay half-life of $^{71}$Zn$^{m}$ could be determined with significantly improved accuracy.
The 
consequences of these data have been studied by
network calculations for convective core He burning and
convective shell C burning in massive stars.
\end{abstract}


\maketitle

\section{Introduction}

The signature of the $s$-process contribution to the solar abundances
implies two parts, a $main$ component, which is responsible for the mass 
region from Y to Bi, and a $weak$ component, which dominates in the region
between Fe and Sr. The main component can be assigned to low mass 
stars with $1 \leq M/M_{\odot} \leq 3$, whereas the weak component
is related to massive stars with $M \geq 8M_{\odot}$ ($M_{\odot}$ 
stands for the mass of the sun). Because of the much shorter evolution 
times of massive stars, the $s$-process enrichment of the Universe starts 
with the lighter $s$ elements. Accordingly, the weak $s$ process is 
important for the early stellar populations and for Galactic chemical 
evolution in general \cite{TGG99,TGA04}. 

The $s$ process in massive stars occurs in two steps during 
different evolutionary phases. Neutrons are mainly produced by the
$^{22}$Ne($\alpha, n$)$^{25}$Mg reaction in both cases, but at
rather different temperatures and neutron densities. During core He
burning, neutrons are produced near core He exhaustion at
temperatures of $T$=(2.5 - 3.5)$\times10^8$ K for about 10$^4$ years
with neutron densities $\lesssim 10^6$ cm$^{-3}$, whereas the higher
temperatures of $T$=(1.0 - 1.4)$\times10^9$ K during the subsequent
carbon shell burning phase give rise to peak neutron densities of about
$10^{12}$ cm$^{-3}$  \cite{RBG91b,RGB93,LSC00}.

In the scenarios for the main and weak $s$ process, the stellar 
($n, \gamma$) cross sections of the involved isotopes constitute 
the essential nuclear physics input, but with an important difference: 
The high neutron exposure during the main component leads to
equilibrium in the reaction flow, expressed by the so-called local
approximation
$ \langle \sigma \rangle N_s  = {\rm constant}$,
which holds true for isotopes between magic neutron numbers. This
relation implies that the emerging $s$ abundances are inversely
proportional to the stellar cross sections and that the uncertainty
of a particular cross section affects only the abundance of that
specific isotope. In contrast, the neutron exposure in massive stars
is too small to achieve flow equilibrium, and this means that cross
section uncertainties are not only influencing the abundance of that
particular isotope but have a potentially strong propagating effect
on the abundances of the subsequent isotopes involved in the
$s$-process chain.

This propagation effect was first discussed for the $^{62}$Ni($n,
\gamma$)$^{63}$Ni reaction \cite{RHH02} and later investigated 
for the complete reaction chain of the weak $s$ process \cite{HKU08a,
HKU08b, PGH10}. Computations with different choices for the Maxwellian
averaged cross sections (MACS) showed important bottle-neck effects 
and the significant impact of crucial capture rates near the Fe seed
on the abundances of the subsequent isotopes in the reaction chain. 
Limitations in the quality of stellar cross sections can, therefore,
have serious consequences for the contributions of the weak $s$
process to Galactic chemical evolution.

Stellar neutron capture cross sections in the mass region of the weak 
$s$ process are often not available with the required accuracy. In 
case of the Zn isotopes, experimental data suffer from large uncertainties 
or are even missing as for $^{70}$Zn \cite{BBK00,DPK09}. Therefore,
a series of activation measurements has been performed 
at the Karlsruhe 3.7 MV Van de Graaff
on natural Zn
to improve the accessible cross sections for $^{64,70}$Zn and for the 
partial cross section of $^{68}$Zn. Because these isotopes have relatively 
small cross sections of less than about 50 mb, they are expected
to give rise to large propagation effects in the final abundance distribution.

Apart from its propagation effect in the weak $s$ process, Zn is of 
general interest for Galactic evolution. Though it is commonly
accepted that Zn is produced by a variety of scenarios 
the respective nucleosynthesis 
mechanisms and their relative contributions are poorly understood. 
Spectroscopic observations over a range of stellar metallicities (e.g. Refs. 
\cite{BIR04,CNZ04,NCA04}) found indications of an overabundance of Zn
compared to Fe in low-Z stars, but an explanation for this excess 
appears premature in view of the present uncertainties concerning the 
origin of Zn. In massive stars, the $s$ process component of the Zn isotopes 
is obscured by the fact that measured neutron capture cross sections are 
scarce, uncertain or even missing.   
 
Experimental technique and measurements are described in Sec. \ref{exp}, 
Secs. \ref{ana} and \ref{resu} deal with data analysis and results. 
Maxwellian average cross sections (MACS) and the related
astrophysical aspects are discussed in Sec. \ref{astro}.
 
\clearpage
\section{Measurements \label{exp}}

\subsection{Experimental technique}

The activation method represents a well established and accurate
approach to determine MACSs at $kT = 25$ keV by producing a
quasi-stellar neutron spectrum via the $^7$Li($p, n$)$^7$Be reaction
\cite{BeK80}. This method has been extensively used, mostly for
measurements related to the main $s$-process component (for
examples see Refs.~\cite{PAK04,BDH03,RAK04}). In the present
experiment, the proton beam with an energy of $E_p$=1912 keV, 30
keV above the reaction threshold, was delivered by the Karlsruhe 3.7
MV Van de Graaff accelerator with typical intensities of 100 $\mu$A.
The neutron production targets consisted of 30 $\mu$m thick metallic
Li layers evaporated onto water cooled copper backings. In this way,
neutrons are kinematically collimated into a forward cone of 120$^\circ$
opening angle. Neutron moderation is avoided since cooling is
achieved by lateral heat conduction to the water flow outside of
this cone. Throughout the irradiations the neutron flux was
continuously monitored and recorded in time steps of typically 60 s
by means of a $^{6}$Li-glass detector at 1~m distance from the
target. This information is important to account for fluctuations of
the neutron yield in evaluating the fraction of the reaction
products that decay already during the irradiations.

For a sketch of the experimental setup and more details see 
Ref.~\cite{HKU08a} for example.

\subsection{Samples and irradiations}

In total, four activations were performed with three metallic Zn 
samples 8 and 12 mm in diameter and 0.5 and 1~mm in thickness 
to minimize uncertainties stemming from sample geometry
(Table~\ref{tab:samples}). The samples were sandwiched between 
0.03 mm thick gold foils of the same diameter, which served as 
neutron flux monitors using the well known Au reference cross 
section \cite{RaK88}. During the activations the samples were 
placed completely inside the neutron cone 
in contact with the neutron target at the position of highest flux. The relative
isotope abundance ratios adopted from Ref.~\cite{DBD03} are
48.268(214) for $^{64}$Zn, 19.024(82) for $^{68}$Zn and 0.631(6) for
$^{70}$Zn.

\begin{table}[htb]
\caption{Zn samples and irradiation parameters.}
\label{tab:samples}
\renewcommand{\arraystretch}{1.5} 
\begin{ruledtabular}
\begin{tabular}{@{}cccccc}
 Activation & Sample & Mass 	& Diameter	& Irradiation & Integrated\\
            &        & (mg)  	& (mm) 	    & time (min)  & flux ($\times 10^{13}$)\\
\hline
 I 	    	& Zn-1  	 & 409.19 	& 12  		& 174  		& 1.12 (4)  \\
 II	    	& Zn-1  	 & 409.19 	& 12		& 2640      & 14.2 (5)  \\
 III	 	& Zn-2  	 & 181.71	& 8  		& 1182		& 7.4 (3)   \\
 IV	 		& Zn-3		 & 800.72	& 12  		& 1242		& 4.9 (3)   \\
\end{tabular}
\end{ruledtabular}
\end{table}

Activation times were chosen between 3 and 44 h according to the 
half lives of the various product nuclei and to test the respective 
time-dependent corrections applied in data analysis. The integrated 
flux values listed in Table \ref{tab:samples} correspond to average 
fluxes between 0.6 and $1.1\times 10^9$ s$^{-1}$, depending on the 
performance of the accelerator and of the Li target.

\subsection{Induced activities}

The induced activities are characterized by energetic $\gamma$-ray
lines. The corresponding decay data are listed in Table \ref{tab:decay}.
 
The higher activities of the gold foils were measured with a 
well calibrated, passively shielded 76~cm$^3$ HPGe detector with 1.7~keV 
resolution at 1.33~MeV $\gamma$-ray energy and a relative efficiency 
of 30\%. 

\begin{figure}[h]
\begin{center}
\includegraphics[width=.45\textwidth]{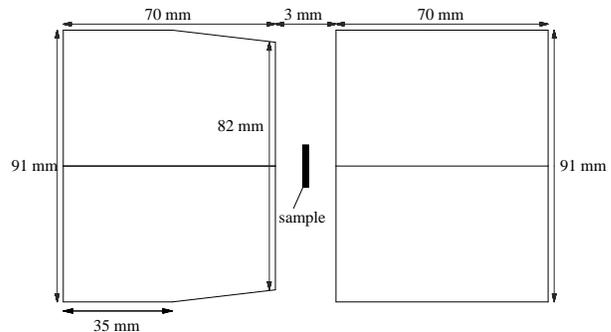}
\caption{Schematic view of the HPGe spectrometer consisting of two
Clover-type detectors in close geometry.}\label{fig:clover}
\end{center}
\end{figure}

The Zn activities were counted with two Clover type HPGe detectors 
facing each other in close geometry. Each Clover detector (Eurisys 
Measures) consists of four independent $n$-type Ge crystals in a 
four-leaf clover arrangement with 0.2~mm gaps in between. The 
originally cylindrical crystals 50~mm in diameter and 70~mm in 
length are shaped as shown in Fig.~\ref{fig:clover}, leaving an 
active volume of about 145~cm$^{3}$ per crystal. The crystals are 
held from the rear through a steel rod 1~mm in diameter and about 35~mm 
in length and are enclosed in a common cryostat. The front end of 
the crystals is separated by a gap of 5~mm from the 1~mm thick aluminum 
window. 

The individual Ge crystals have a resolution of typically 2.0~keV at 
1.33~MeV. At 122~keV, the peak-to-total ratio is $\sim$45.
The detector can be operated either in single mode by considering 
the signals from each crystal independently or in calorimetric mode, 
when coincident signals from different crystals are added off-line. In 
this way Compton-scattered events can be restored in the sum spectrum 
if the scattered photon is detected in one of the neighbouring crystals,
resulting in a significantly higher full-energy-peak efficiency. 

In both setups, the counting position could be reproduced within 
0.1 mm by means of special adapters. This feature was crucial for
obtaining a well defined efficiency, especially for the close 
geometry of the two Clover detectors.

\begin{figure}
\includegraphics[width=20pc]{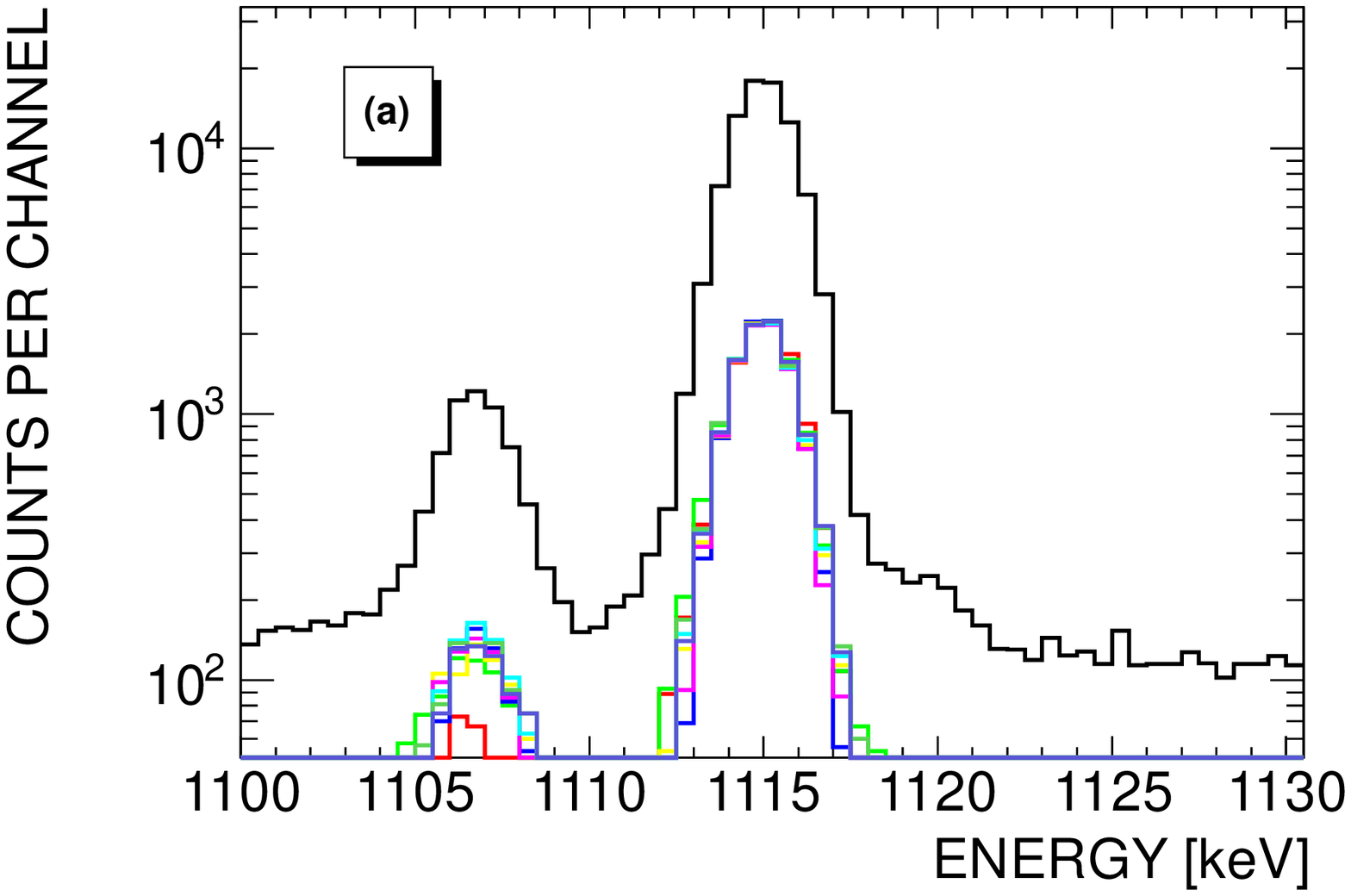}
\includegraphics[width=20pc]{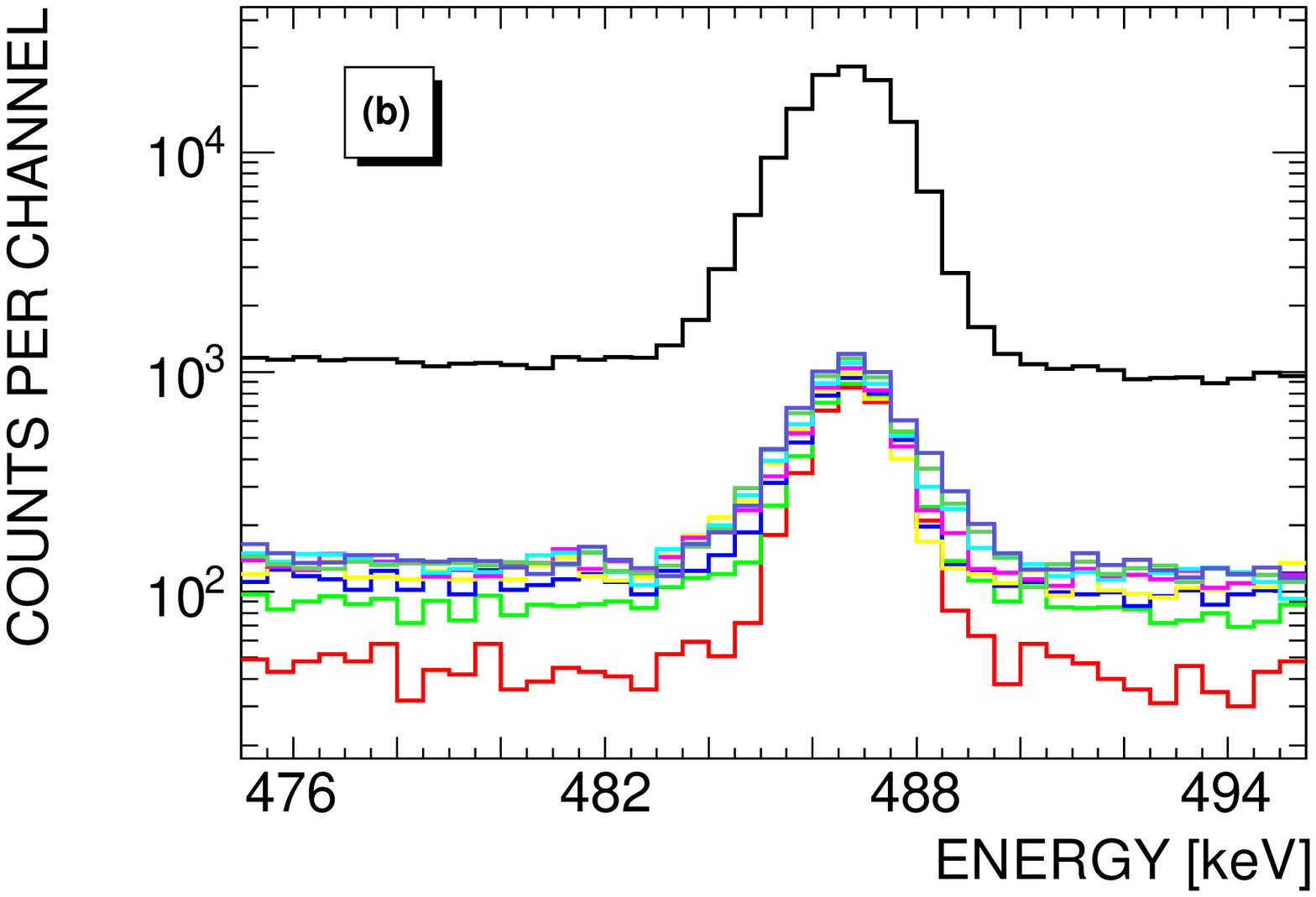}
\includegraphics[width=20pc]{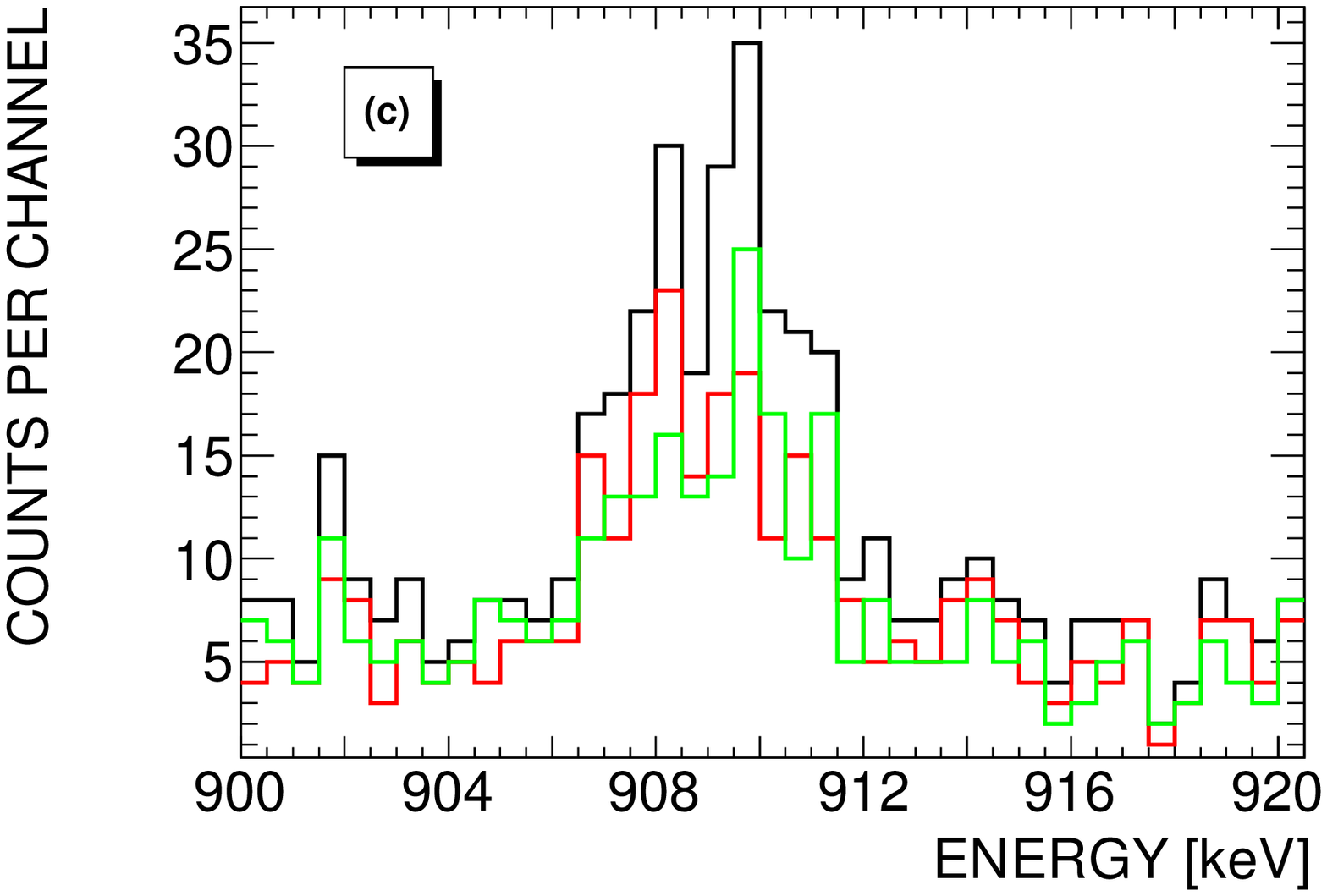}
\caption{(Color online) Examples for the $\gamma$-ray spectra measured after 
the irradiations. Shown are from top to bottom the decay lines of $^{65}$Zn
at 1115 keV (a), $^{71}$Zn$^{\rm m}$ at 487 keV (b), and the very weak
line from $^{71}$Zn$^{\rm g}$ at 910 keV (c). The spectra of the eight 
individual crystals in the HPGe Clover detectors are plotted together with 
the respective sum. The calorimetric mode was used only in case (c) for 
improving the signal/background ratio.  
\label{fig:exspec}}
\end{figure}

Examples for the accumulated $\gamma$ spectra are shown in Fig. 
\ref{fig:exspec} for the 1115~keV line in the decay of $^{65}$Zn, 
for the 487~keV line from $^{71}$Zn$^{\rm m}$ and
for the 910~keV line from $^{71}$Zn$^{\rm g}$.
.

\begin{table}
\caption{Decay properties of the product nuclei.}
\label{tab:decay}
\renewcommand{\arraystretch}{1.5} 
\begin{ruledtabular}
\begin{tabular}{@{}ccccc}
 Product     		 & Half-life  & $\gamma$-Ray energy & Intensity  & Ref.\\
 nucleus           	 & 			  & (keV)     	 		& per decay (\%) & 	 	    \\
\hline
$^{65}$Zn   		 & 243.93(9) d   & 1115.539(2)  	& 50.04(10) & \cite{BrT10} \\
$^{69}$Zn$^{\rm m}$  & 13.76(2) h    & 438.636(18)	& 94.8(2)   & \cite{BhT00} \\
$^{71}$Zn$^{\rm g}$  & 2.45(10) min  & 910.3(1)    	& 7.84(64)  & \cite{AbS11} \\
$^{71}$Zn$^{\rm m}$  & 3.96(5) h     & 487.30(4)    	& 61.2(23)  & \cite{AbS11} \\
$^{198}$Au    	     & 2.69517 (21) d & 411.80205 (17)	& 95.62(6)  & \cite{Hua09}	\\
\end{tabular}
\end{ruledtabular}
\end{table}

\begin{figure}[h]
\begin{center}
\includegraphics[width=.45\textwidth]{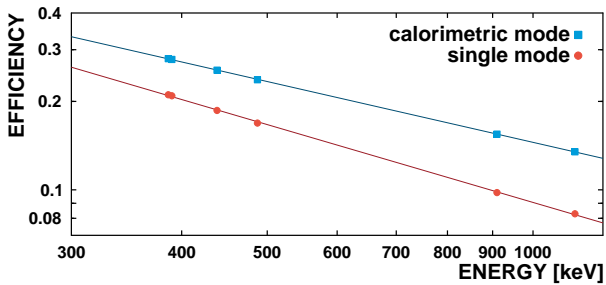}
\caption{(Color online) Comparison of the $\gamma$-ray efficiency of
the Clover setup for the two detection modes. The symbols show simulated
efficiencies for the $\gamma$-rays emitted in the decay of $^{65}$Zn,
$^{69}$Zn$^{\rm m}$, $^{71}$Zn$^{\rm m}$ and $^{71}$Zn$^{\rm g}$,
respectively. The lines represent a fit of the function $\varepsilon
(E) = a \cdot E^{b}$ to the simulated values. The efficiency in
calorimetric mode is on average a factor of 1.45 increased compared to the
single mode.}\label{fig:clov-eff}
\end{center}
\end{figure}

The efficiency of the Clover setup was determined with a set of 
point-like calibration sources 
($^{22}$Na, $^{54}$Mn, $^{60}$Co, $^{65}$Zn, $^{88}$Y, $^{131}$Ba, $^{137}$Cs)
complemented by a detailed Monte-Carlo simulation using a full model of the 
setup as sketched above. These simulations were particularly
important for determining the efficiencies in the calorimetric mode, 
which was used for the 910 keV transition in the decay 
of $^{71}$Zn$^{\rm g}$. The absolute efficiencies for the two detection 
modes of the Clover setup are compared in Fig.~\ref{fig:clov-eff}. A 
detailed description of the simulations and the calibration procedure 
is given in Ref.~\cite{DPA04}. 

The simulations provided also the small corrections to  the measured
peak detection efficiency values for the point-like calibration sources 
($\epsilon_\gamma$). These corrections, which are listed in 
Table~\ref{tab:corrections}, refer to the actual diameter of the samples 
 (K$_{ext}$), to $\gamma$-ray absorption in the sample (K$_{abs}$), 
and to the possible summing of cascade transitions (K$_{sum}$).
 
\begin{table}[htb]
\caption{Efficiencies and correction factors compared to ideal point-like sources 
for the Clover setup
in single mode. Only the data for the 910~keV line of $^{71}$Zn$^{\rm g}$ are treated 
in calorimetric mode.}
\label{tab:corrections}
\renewcommand{\arraystretch}{1.5} 
\begin{ruledtabular}
\begin{tabular}{@{}ccccccc}
Product 		& E$_\gamma$ & $\epsilon_\gamma$ & Sample & K$_{ext}$ & K$_{abs}$  & K$_{sum}$ \\
nucleus         & (keV) 	 &         		     &        & 	      & 		   &           \\
\hline
$^{65}$Zn  		   & 1115 	 & 0.0829 		     & Zn-1   & 0.9894 	  & 0.9877 	   & 1.0011    \\
 			       &  		 &  		   	     & Zn-2   & 0.9956 	  & 0.9836 	   & 1.0006    \\
 			       &  		 &  			     & Zn-3   & 0.9884 	  & 0.9772 	   & 0.9978    \\
$^{69}$Zn$^{\rm m}$& 439     & 0.1863 		     & Zn-1   & 0.9901 	  & 0.9794 	   & 0.999     \\
 			       &  		 &  			     & Zn-2   & 0.9938 	  & 0.9783	   & 0.9991    \\
 			       &  		 &  			     & Zn-3   & 0.9906	  & 0.9592 	   & 0.9994    \\
$^{71}$Zn$^{\rm m}$& 487     & 0.1688 		     & Zn-1   & 0.9907 	  & 0.9808 	   & 0.8577    \\
 			       &  		 &  			     & Zn-2   & 0.9949 	  & 0.9808 	   & 0.8655    \\
 			       &	   	 &  			     & Zn-3   & 0.9924 	  & 0.9606 	   & 0.8536    \\
$^{71}$Zn$^{\rm g}$& 910     & 0.1547		     & Zn-1   & 0.9955 	  & 0.9847 	   & 0.963 	   \\
\end{tabular}
\end{ruledtabular}
\end{table}

\clearpage
\section{Data analysis \label{ana}}

\subsection{Half-life of $^{71}$Zn$^{\rm m}$}

The induced activities were followed in time steps of 30~min to
exclude parasitic activities, which might have contributed to the
characteristic $\gamma$ lines listed in Table \ref{tab:decay}.
The time sequence obtained for $^{71}$Zn$^{\rm m}$
showed a significant deviation from what was expected from the recommended
half life of $t_{1/2}^{rec} = 3.96(5)$~h \cite{AbS11}. This
value represents a weighted average of measurements, which were
performed about 50 years ago using different techniques to produce
$^{71}$Zn$^{\rm m}$ and observe its decay \cite{Lev58,ThP61,SEA64}
(see Table \ref{tab:halflife}).

\begin{table}[htb]
\caption{Half-life of $^{71}$Zn$^{\rm m}$. The two left columns
summarize the values of previous measurements and their weighted
average, the recommended value by \cite{AbS11}. In comparison, the two
right columns show the results obtained in this work from the decay
curves of three $\gamma$ transitions in $^{71}$Ga and their weighted
average (w.a.). The uncertainties correspond to one standard
deviation.}
\label{tab:halflife}
\begin{ruledtabular}
\begin{tabular}{llll}
\multicolumn{2}{c}{literature values} &
\multicolumn{2}{c}{this work} \\
Ref.         & $t_{1/2}$ (h) & E$_{\gamma}$ (keV) & $t_{1/2}$ (h)\\
\hline
\cite{Lev58} & 3.92(5)       & 386                & 4.142(9)     \\
\cite{ThP61} & 4.0(1)        & 487                & 4.117(12)    \\
\cite{SEA64} & 4.1(1)        & 620                & 4.098(15)    \\
w.a.~\cite{AbS11} & 3.96(5)       & w.a.               & 4.125(7)     \\
\end{tabular}
\end{ruledtabular}
\end{table}

\begin{figure}[h]
\begin{center}
\includegraphics[width=.45\textwidth]{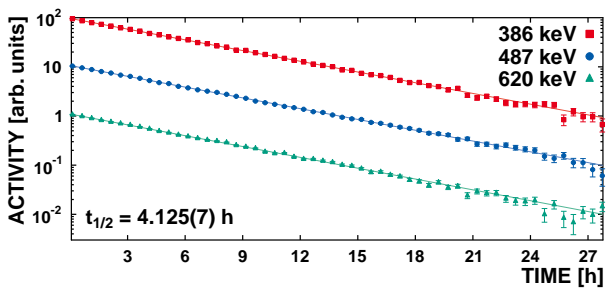}
\caption{(Color online) Measured decay curves of $^{71}$Zn$^{\rm m}$ for
three $\gamma$ transitions in $^{71}$Ga including a fit to the data, see Table~\ref{tab:halflife}.
The error bars are smaller than the symbols. The
given value of $t_{1/2} = 4.125(7)$ h is the weighted average derived
from the shown decay curves. It differs significantly from the
previously recommended half-life of 3.96(5)~h \cite{AbS11}.}
\label{fig:decay}
\end{center}
\end{figure}

The decay curves based on the $\gamma$ transitions with energies of
386, 487, and 620 keV in $^{71}$Ga are shown in Fig. \ref{fig:decay}
for about seven half-lives of $^{71}$Zn$^{\rm m}$. The activity was
followed for more than ten half-lives until no significant signals
above background could be detected. The weighted average derived from
a fit to these decay curves results in new half life of 4.125(7)~h
with a significantly improved uncertainty compared to the previously
recommended value \cite{AbS11}.

Combining this new result with the former values as listed in
Tab.~\ref{tab:halflife} yields a weighted average of 4.120(7)~h. Only
with this revised half life the deduced cross section for the reaction
$^{71}$Zn$^{\rm m}$ became independent of the time interval chosen for
the activity determination.  This was impossible to achieve using the
previously recommended half life \cite{AbS11}.

\subsection{Cross section determination}

The total number of activated nuclei $A$ is given by

\begin{equation}
A = \phi \, N \, \sigma \, f_{b}
\end{equation}

\noindent where $\phi$ is the time integrated neutron flux, $N$ the
number of sample atoms, and $\sigma$ the spectrum averaged neutron
capture cross section. The factor $f_{b}$ accounts for variations of
the neutron flux and for the decay during activation.

The number of activated nuclei in Eq.(1) is determined from the
number of counts in a characteristic $\gamma$-ray line,

\begin{equation}
C_\gamma=A \, K_\gamma \, \varepsilon _\gamma \,
           I_\gamma \, (1-{\rm exp}(-\lambda t_m))\,
           {\rm exp}(-\lambda t_w)
\end{equation}

\noindent
where $K_\gamma$ combines the correction factors listed in columns 
5 - 7 of Table~\ref{tab:corrections}, $\varepsilon _{\gamma}$ the 
efficiency of the HPGe-detection system, $I_\gamma$ the line intensity, 
$t_w$ the waiting time between irradiation and $\gamma$-spectroscopy, 
and $t_m$ 
the duration of the activity measurement. 

\begin{figure}
\includegraphics[width=20pc]{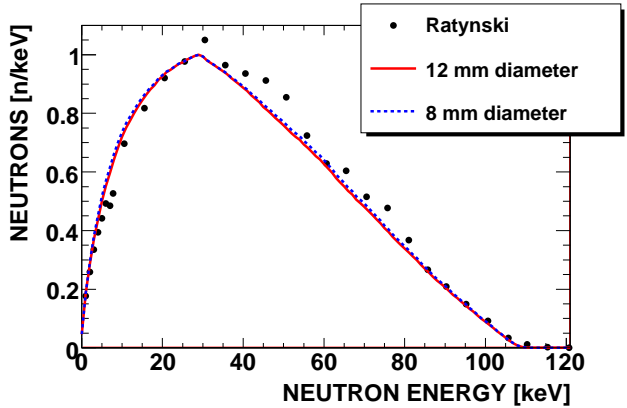}
\caption{(Color online) Simulated neutron spectra for the two sample diameters in 
comparison with the  spectrum of Ref.~\cite{RaK88}. The simulated 
spectra contain a weighting factor for neutrons emitted at larger angles 
with respect to the incoming proton beam, which is not included in the 
experimental spectrum of \cite{RaK88}. Details of the simulations can 
be found in \cite{RHK09}.
\label{fig:simspec}}
\end{figure}

The time-integrated flux at the sample position, 
$\phi$, is determined from the intensities of the 412 keV
$\gamma$-ray line in the spectra of the gold foils \cite{RAH03},
\begin{equation}   
\phi = \frac{\phi_{1}+\phi_{2}}{2}
\end{equation}
with
\begin{equation}   
\Delta\phi  = \frac{\phi_{1}-\phi_{2}}{4}.
\end{equation}
The neutron flux seen by the gold samples follows from the number of
$^{198}$Au nuclei,
\begin{equation} 
\phi = \frac{N_{198}}{N_{197}\,\sigma\,f_b}
\end{equation}
where the correction $f_b$ accounts for the fraction of $^{198}$Au
nuclei that decayed already during the irradiation \cite{BeK80}. The 
spectrum-integrated $^{197}$Au($n,\gamma$) cross section is obtained by 
folding the corresponding neutron spectra, which were calculated with 
the code PINO \cite{RHK09} (Fig.~\ref{fig:simspec}) with the 
differential $^{197}$Au($n,\gamma$) cross section of Macklin \cite{Mac82a} 
normalized to the value for $kT=25$~keV of Ratynski and K{\"a}ppeler 
\cite{RaK88}.

The integrated flux values determined by the gold foils and the 
corresponding average seen by the Zn samples are given in 
Table~\ref{tab:run-parm} for the activation runs I-IV. 

\begin{table}[htb]
\caption{Integrated neutron fluxes (in units of 10$^{13}$) and activation times $t_a$.  }
\renewcommand{\arraystretch}{1.2} 
\label{tab:run-parm}
\begin{ruledtabular}
\begin{tabular}{@{}ccccc}
Run     & $\phi{_{1}}$ & $\phi{_{2}}$ & $\phi{_{sample}}$ & $t_a$ (h) \\
\hline
I       &  1.20        & 1.05         &  1.12 $\pm$ 0.04  & 2.9   	  \\
II      &  15.2        & 13.2    	  &  14.2 $\pm$ 0.5   & 44.0  	  \\
III     &  7.9         & 6.8     	  &  7.4  $\pm$ 0.3   & 19.7  	  \\
IV      &  5.5         & 4.4     	  &  4.9  $\pm$ 0.3   & 20.7  	  \\
\end{tabular}
\end{ruledtabular}
\end{table}

\subsection{Uncertainties}

The experimental uncertainties are summarized in Table
\ref{tab:uncertain}, where the investigated reactions are indicated
by the respective product nuclei.

Significant contributions to the overall uncertainty originate from
the gold reference cross section, the efficiency of the HPGe
detectors, and the time integrated neutron flux. The $^{70}$Zn 
cross sections are also affected by uncertain $\gamma$-decay 
intensities. Any improvement of these data would, therefore, be
important.

Because statistical uncertainties were found to be practically 
negligible after averaging the results from repeated activations,
the final uncertainties are determined by systematic effects. 
Only in case of the partial cross section to the ground 
state of $^{71}$Zn, the counting statistics was not sufficient 
for achieving an adequate accuracy of the final result.

\begin{table}[htb]
\caption{Compilation of uncertainties.
\label{tab:uncertain}}
\begin{ruledtabular}
\begin{tabular} {lccccc}
Source of Uncertainty & \multicolumn{5}{c}{Uncertainty (\%)}  \\
\cline{2-6}
 & Au & $^{65}$Zn & $^{69}$Zn$^{\rm m}$ & $^{71}$Zn$^{\rm g}$ & $^{71}$Zn$^{\rm m}$ \\
\hline
Gold cross section                            & 1.5   &  -    & -     & -    & -    \\
Number of nuclei                              & 0.1   & 0.5   & 0.4   & 0.9  &  0.9 \\
Time factors, $f_i$, $t _{1/2}$ 			  & 	  & 0.1   & 0.2   & 4.0  &  1.3 \\
Self-absorption, $K_\gamma$                   & \multicolumn{5}{c}{0.2} 	        \\
Detector efficiency, $\epsilon_\gamma$        & \multicolumn{5}{c}{1.5}       	    \\
$\gamma$-Ray intensity, $I_\gamma$  		  &  0.06 & 0.2   &  0.2  & 8.2  &  3.8 \\
Time integrated flux, $\phi$          		  & -     & 1.5   & 1.5   & 2.5  &  2.4 \\
Counting statistics                			  &       & 0.7   & 0.9   & 13   &  0.8 \\
                                           	  &       &       &       &      &      \\
Total uncertainty                 		      & -     & 2.8   & 2.8   & 16   &  5.1 \\
\end{tabular}
\end{ruledtabular}
\end{table}

\clearpage
\section{Results \label{resu}}

\subsection{Measured cross sections}

Table \ref{tab:results} shows a summary of the results obtained in
all four activations together with the corresponding uncertainties. In
spite of the variation of the experimental parameters (see Table
\ref{tab:samples}), the results are all consistent within the estimated
uncertainties, thus confirming the procedures applied in data
analysis. These variations included different sample sizes and
masses to verify the corrections for finite size and self shielding
effects as well as different irradiation times to control
uncertainties due to the half-life of the respective product
nucleus.

\begin{table}[htb]
\caption{Activations, $\gamma$-spectroscopy, and cross section results$^a$.}
\label{tab:results}
\begin{ruledtabular}
\begin{tabular}{@{}ccccc}
Activation & $\gamma$-Ray energy & Cross section     & Mean value   \\
           & (keV)               & (mb)              & (mb)         \\
\hline
\multicolumn{4}{c}{$^{64}$Zn($n, \gamma$)$^{65}$Zn}	   				\\
       I   & 1115.5              & 54.5$\pm$2.1      &              \\
  	   II  & 1115.5  			 & 54.7$\pm$2.2  	 & 				\\
	   III & 1115.5  			 & 52.7$\pm$2.2  	 & 				\\
	   IV  & 1115.5  			 & 52.6$\pm$3.0  	 &  53.5$\pm$1.5\\
		   &     	 			 &         			 &              \\
\multicolumn{4}{c}{$^{68}$Zn($n, \gamma$)$^{69}$Zn$^{\rm m}$} 		\\
       I   & 438.64              & 3.27$\pm$0.13     & 				\\
	   II  & 438.64  			 & 3.38$\pm$0.14 	 & 				\\
	   III & 438.64  			 & 3.25$\pm$0.14 	 & 				\\
	   IV  & 438.64  			 & 3.31$\pm$0.19 	 & 3.30$\pm$0.09\\
		   &     				 &         			 &              \\
\multicolumn{4}{c}{$^{70}$Zn($n, \gamma$)$^{71}$Zn$^{\rm g}$} 		\\
       I   & 910.27              & 4.18$\pm$0.67     & 4.18$\pm$0.67$^b$\\
		   &     				 &         			 &              \\
\multicolumn{4}{c}{$^{70}$Zn($n, \gamma$)$^{71}$Zn$^{\rm m}$}		\\
       I   & 487.38              & 6.69$\pm$0.40     &  			\\
	   II  & 487.38  			 & 7.46$\pm$0.43 	 &  			\\
	   III & 487.38  			 & 6.39$\pm$0.37 	 &  			\\
	   IV  & 487.38  			 & 6.75$\pm$0.47 	 & 6.79$\pm$0.34       \\
\end{tabular}\\
\end{ruledtabular}
$^a$Cross section averaged over quasi-stellar spectrum.\\
$^b$Only activation I could be considered, all other activations lasted too long.
\end{table}

Previous ($n, \gamma$) cross section data for the investigated isotopes 
$^{64,68,70}$Zn are rather limited \cite{BBK00,DHK06}. For $^{64}$Zn 
as well as for $^{68}$Zn, there is only a single measurement based on 
the time-of-flight (TOF) method. In both cases, uncertainties of 10\% have 
been reported \cite{GTH81,GTH82}. The partial cross section to the isomer 
in $^{69}$Zn has been studied once \cite{MSR73} and was quoted with 
an uncertainty of 25\%, whereas there are no experimental results for 
$^{70}$Zn at all. 

As illustrated in Table \ref{tab:results}, these uncertainties could be 
substantially improved. The numerical cross section data are compared for 
the MACS values in the following section.

\subsection{Maxwellian averages}

The values listed in Table \ref{tab:results} represent average cross 
sections for the experimental neutron distribution used in the 
irradiations. Though the experimental spectrum corresponds in very 
good approximation to a thermal spectrum for $kT=25$ keV, the cut-off 
at 106\,keV requires a small correction, in particular if the 
investigated cross section exhibits a different energy  dependence 
than the gold reference cross section. 

For the calculation of the final MACSs
\begin{equation}
\label{macs}
\langle\sigma\rangle_{kT}=\frac{\langle\sigma v\rangle}{v_T}=\frac{2}{\sqrt{\pi}}
              \frac{\int_0^{\infty} \sigma(E_n)\, E_n \,
              {\rm exp}(-E_n/kT)\, dE_n}{\int_0^{\infty} E_n \,
              {\rm exp}(-E_n/kT)\, dE_n}
\end{equation}
\noindent
the correction was obtained by normalizing the differential ($n, 
\gamma$) cross sections, $\sigma_{n,\gamma}$($E_n$), e.g. from 
theoretical calculations with the Hauser-Feshbach (HF) statistical model
\cite{RaT00} or from evaluated data libraries (http://www-nds.iaea.org/),
with the new experimental values. Apart from the normalization
factor $2/\sqrt{\pi}$ from the definition of the MACS in Eq. 
\ref{macs}, these corrections are between 3\% and 5\%. 

The normalized energy-dependent cross sections can also be used 
for extrapolation to other temperatures as shown in Table 
\ref{tab:macs}. The data from the compilation of 
Refs.~\cite{BBK00,DPK06} are listed before and after normalization
to the present results for $kT=25$~keV

\begin{table*}[htb]
\caption{MACSs of $^{64}$Zn and $^{70}$Zn compared to
the compilation of Bao {\it et al.} \cite{BBK00}.$^a$ 
Since only the partial cross section of $^{68}$Zn($n,\gamma$$^{69}$Zn$^{\mbox{m}}$ 
was measured here, 
only the total MACS of $^{68}$Zn from \cite{BBK00} is given.}
\label{tab:macs}
\begin{ruledtabular}
\begin{tabular}{@{}lccccccccccc}
     &    \multicolumn{11}{c}{MACS (mb)}  \\
\cline{2-12}
$kT$ (keV)        & 5    & 10   & 15   & 20   & 25   & 30       & 40  & 50  & 60  & 80  & 100  \\
\hline
          		  &  	 &     &     &     &       &              &      &     &     &     &      \\
\multicolumn{11}{c}{$^{64}$Zn($n, \gamma$)$^{65}$Zn} \\
Ref.~\cite{BBK00}     & 139  & 108 &  88 &  75 &  66   & 59$\pm$5     & 52   & 47  & 44  & 40  & 38   \\
Ref.~\cite{BBK00}$^*$ & 123  & 95.9& 78.2& 66.6&  58.6$\pm$1.7 & 52.4$\pm$1.7 & 46.2 & 41.7& 39.1& 35.5& 33.7 \\
          		  &  	 &     &     &     &       &              &      &     &     &     &      \\
\multicolumn{11}{c}{$^{68}$Zn($n, \gamma$)$^{69}$Zn} \\
Ref.~\cite{BBK00}     & 331  & 238 & 197 & 174 & 153   &139$\pm$6     &121   & 113 & 102 & 87  & 79  \\
          		  &  	 &     &     &     &       &              &      &     &     &     &      \\
\multicolumn{11}{c}{$^{70}$Zn($n, \gamma$)$^{71}$Zn} \\
Ref.~\cite{BBK00}     & 57   & 38  & 30  & 26.3& 23.5  & 21.5$\pm$2   & 18.7 & 16.9& 15.5& 13.7& 12.5 \\
Ref.~\cite{BBK00}$^*$ & 28.4 & 18.9& 14.9& 13.1& 11.7$\pm$0.8  & 10.7$\pm$0.8 & 9.3  & 8.4 &  7.7& 6.8 & 6.2  \\
\end{tabular}
\end{ruledtabular}
$^a$ For comparison with measured cross section values in Table \ref{tab:results}
multiply with $\sqrt{\pi}/2$ \\
$^*$ Normalized to the measured value.
\end{table*}

MACS values are commonly compared at the standard thermal energy of
$kT=30$ keV (Table \ref{tab:macs}). The value for $^{64}$Zn is about 
12\% lower than the MACS based on previous TOF data \cite{GTH81}, but
the uncertainty could be reduced by a factor of 2.5. In case 
of $^{70}$Zn, the sum of the partial cross sections is considerably 
smaller than the previously compiled value that had been obtained by
an HF calculation and an additional empirical correction factor
\cite{BBK00}. The factor of two difference found for the $^{70}$Zn 
cross section, which had been inferred from purely theoretical HF predictions 
\cite{RaT00,Har81,WFH78,Gor02,Gor05}, is reflecting the uncertainty of 
the HF approach in this mass region.

\clearpage
\section{Astrophysical implications \label{astro}}
\begin{figure}[h]
\begin{center}
\includegraphics[width=.45\textwidth]{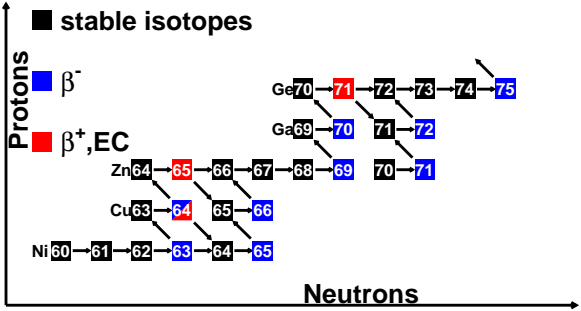}
\caption{(Color online) $s$-process reaction network in the region around Zn. }
\label{fig:network}
\end{center}
\end{figure}

Fig.~\ref{fig:network} shows the $s$-process path in the region around zinc.
There are two potential branchings in the reaction flow affecting the production
of the important $s$-only nucleus $^{70}$Ge. The first branching bypassing  $^{70}$Ge
could occur between $\beta^{-}$-decay and neutron capture at $^{69}$Zn. 
However this branching is not open, since 
the $\beta^{-}$-halflives of ground and isomeric state are too short
(1~h and 14~h). Therefore, $^{69}$Zn will always undergo a $\beta^{-}$-decay
before capturing a neutron.
The second potential branching could occur
between $\beta^{-}$- and EC-decays at $^{70}$Ga. 
However, the EC-decays
occurs only in 0.4\% of all decays, hence the branching can be neglected.
This means, the s-process path is not even partly bypassing the s-only 
nucleus $^{70}$Ge, hence $^{70}$Ge is a well suited normalization point between
s-abundance and solar abundance. On the other hand, the $s$-process path is 
completely bypassing $^{70}$Zn, which makes it
an $r$-only nucleus.

The $s$ process in the zinc region was investigated using 
the nucleosynthesis code {\small NETZ} \cite{Jaa91}. The 
information on the respective stellar 
scenarios, i.e. the effective time dependent profiles for 
temperature, mass density, and neutron density in the considered
burning zones were adopted from \cite{RGB93,RBG91a}. The cross sections used 
are the latest recommended values from the {\small KADoNiS} database \cite{DHK06},
which followed the compilation of \cite{BBK00}.

\begin{figure}[h]
\begin{center}
\includegraphics[width=.45\textwidth]{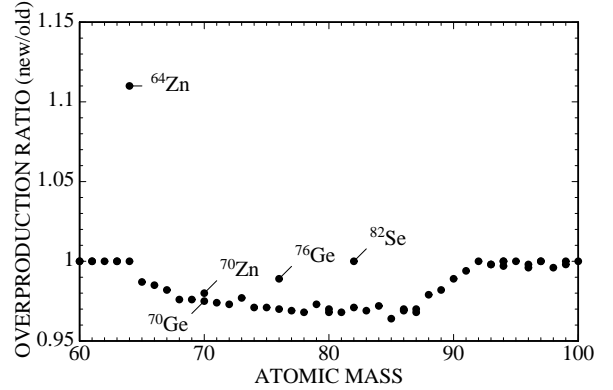}
\caption{The effect of the new $^{64}$Zn for the overproduction in the weak $s$ process. 
$^{70}$Ge is used as a normalization 
point of the $s$-process abundances. Because $^{70}$Zn is bypassed during the
weak $s$-process, its strongly reduced cross section has no significant impact. 
This holds true also for the neutron rich isotopes of germanium
and selenium ($^{76}$Ge, $^{82}$Se).}\label{fig:sensitivity}
\end{center}
\end{figure}

The most important change in the abundance distribution of the weak $s$ process results 
from the 
new, reduced neutron capture cross section of $^{64}$Zn. The comparison of the 
overabundance factors in Fig.~\ref{fig:sensitivity} 
shows that the abundance of $^{64}$Zn is increased by 11\%,
while most of the heavier 
elements up until mass 90 are slightly less produced because of 
the aforementioned propagation effect that is typical for the 
conditions of the weak $s$ process. This effect reflects 
the fact that the local 
equilibrium of the s-process is not reached during the conditions of the weak
$s$-process component.

\clearpage \section{Summary} The neutron capture cross sections of
$^{64}$Zn, $^{68}$Zn, and $^{70}$Zn have been measured with the
activation technique in a quasi-stellar neutron spectrum corresponding
to a thermal energy of $kT = 25$ keV. Samples of natural isotopic
composition have been used.

By a series of repeated irradiations with different experimental
conditions, the $^{64}$Zn($n,\gamma$)$^{65}$Zn cross section was
determined to 53.5$\pm$1.5~mb and the partial cross section
$^{68}$Zn($n,\gamma$)$^{69}$Zn$^{m}$ feeding the isomeric state in
$^{69}$Zn to 3.30$\pm$0.09~mb for the experimental spectrum.  Values
of 6.79$\pm$0.34~mb and 4.2$\pm$0.7~mb were found for the partial
cross sections of $^{70}$Zn($n,\gamma$)$^{71}$Zn$^{m}$ and
$^{70}$Zn($n,\gamma$)$^{71}$Zn$^{g}$, which had not been measured so
far.

The half-life of $^{71}$Zn$^{\rm m}$ could be determined to
4.125$\pm$0.007~h, with a significantly improved accuracy compared to
the previous recommendation. The weighted average of the combined data
yields a value of 4.120$\pm$0.007~h.

Compared to previous measurements on $^{64,68}$Zn, the uncertainties
could also be significantly improved, while the $^{70}$Zn cross
section was found two times smaller than existing model
calculations. From these results Maxwellian average cross sections
were determined between 5 and 100~keV.

The consequences of these data have been studied by network
calculations based on the temperature and neutron density profiles for
convective core He burning and convective shell C burning in massive
stars.  These simulations of the weak $s$ process reveiled that only
the change of the $^{64}$Zn($n,\gamma$)$^{65}$Zn cross section results
in significant abundance changes compared to the previous set of MACS
values. The other cross sections have much smaller impact because the
MACS of $^{68}$Zn did not change and $^{70}$Zn lies outside the main
path of the weak $s$-process component.

\begin{acknowledgments}
We would like to thank E.-P.~Knaetsch, D.~Roller, and W.~Seith for their
support at the Karlsruhe Van de Graaff accelerator. This work was 
supported by the HGF Young Investigators Project VH-NG-327 and the 
EuroGenesis project MASCHE.

\end{acknowledgments}

\clearpage

\newcommand{\noopsort}[1]{} \newcommand{\printfirst}[2]{#1}
  \newcommand{\singleletter}[1]{#1} \newcommand{\swithchargs}[2]{#2#1}


\begin{thebibliography}{43}
\expandafter\ifx\csname natexlab\endcsname\relax\def\natexlab#1{#1}\fi
\expandafter\ifx\csname bibnamefont\endcsname\relax
  \def\bibnamefont#1{#1}\fi
\expandafter\ifx\csname bibfnamefont\endcsname\relax
  \def\bibfnamefont#1{#1}\fi
\expandafter\ifx\csname citenamefont\endcsname\relax
  \def\citenamefont#1{#1}\fi
\expandafter\ifx\csname url\endcsname\relax
  \def\url#1{\texttt{#1}}\fi
\expandafter\ifx\csname urlprefix\endcsname\relax\def\urlprefix{URL }\fi
\providecommand{\bibinfo}[2]{#2}
\providecommand{\eprint}[2][]{\url{#2}}

\bibitem[{\citenamefont{Travaglio et~al.}(1999)\citenamefont{Travaglio, Galli,
  Gallino, Busso, Ferrini, and Straniero}}]{TGG99}
\bibinfo{author}{\bibfnamefont{C.}~\bibnamefont{Travaglio}},
  \bibinfo{author}{\bibfnamefont{D.}~\bibnamefont{Galli}},
  \bibinfo{author}{\bibfnamefont{R.}~\bibnamefont{Gallino}},
  \bibinfo{author}{\bibfnamefont{M.}~\bibnamefont{Busso}},
  \bibinfo{author}{\bibfnamefont{F.}~\bibnamefont{Ferrini}}, \bibnamefont{and}
  \bibinfo{author}{\bibfnamefont{O.}~\bibnamefont{Straniero}},
  \bibinfo{journal}{Ap. J.} \textbf{\bibinfo{volume}{521}},
  \bibinfo{pages}{691} (\bibinfo{year}{1999}).

\bibitem[{\citenamefont{Travaglio et~al.}(2004)\citenamefont{Travaglio,
  Gallino, Arnone, Cowan, Jordan, and Sneden}}]{TGA04}
\bibinfo{author}{\bibfnamefont{C.}~\bibnamefont{Travaglio}},
  \bibinfo{author}{\bibfnamefont{R.}~\bibnamefont{Gallino}},
  \bibinfo{author}{\bibfnamefont{E.}~\bibnamefont{Arnone}},
  \bibinfo{author}{\bibfnamefont{J.}~\bibnamefont{Cowan}},
  \bibinfo{author}{\bibfnamefont{F.}~\bibnamefont{Jordan}}, \bibnamefont{and}
  \bibinfo{author}{\bibfnamefont{C.}~\bibnamefont{Sneden}},
  \bibinfo{journal}{Ap. J.} \textbf{\bibinfo{volume}{601}},
  \bibinfo{pages}{864} (\bibinfo{year}{2004}).

\bibitem[{\citenamefont{Raiteri
  et~al.}(1991{\natexlab{a}})\citenamefont{Raiteri, Busso, Gallino, and
  Picchio}}]{RBG91b}
\bibinfo{author}{\bibfnamefont{C.}~\bibnamefont{Raiteri}},
  \bibinfo{author}{\bibfnamefont{M.}~\bibnamefont{Busso}},
  \bibinfo{author}{\bibfnamefont{R.}~\bibnamefont{Gallino}}, \bibnamefont{and}
  \bibinfo{author}{\bibfnamefont{G.}~\bibnamefont{Picchio}},
  \bibinfo{journal}{Ap. J.} \textbf{\bibinfo{volume}{371}},
  \bibinfo{pages}{665} (\bibinfo{year}{1991}{\natexlab{a}}).

\bibitem[{\citenamefont{Raiteri et~al.}(1993)\citenamefont{Raiteri, Gallino,
  Busso, Neuberger, and K{\"a}ppeler}}]{RGB93}
\bibinfo{author}{\bibfnamefont{C.}~\bibnamefont{Raiteri}},
  \bibinfo{author}{\bibfnamefont{R.}~\bibnamefont{Gallino}},
  \bibinfo{author}{\bibfnamefont{M.}~\bibnamefont{Busso}},
  \bibinfo{author}{\bibfnamefont{D.}~\bibnamefont{Neuberger}},
  \bibnamefont{and}
  \bibinfo{author}{\bibfnamefont{F.}~\bibnamefont{K{\"a}ppeler}},
  \bibinfo{journal}{Ap. J.} \textbf{\bibinfo{volume}{419}},
  \bibinfo{pages}{207} (\bibinfo{year}{1993}).

\bibitem[{\citenamefont{Limongi et~al.}(2000)\citenamefont{Limongi, Straniero,
  and Chieffi}}]{LSC00}
\bibinfo{author}{\bibfnamefont{M.}~\bibnamefont{Limongi}},
  \bibinfo{author}{\bibfnamefont{O.}~\bibnamefont{Straniero}},
  \bibnamefont{and} \bibinfo{author}{\bibfnamefont{A.}~\bibnamefont{Chieffi}},
  \bibinfo{journal}{Ap. J. Suppl.} \textbf{\bibinfo{volume}{129}},
  \bibinfo{pages}{625} (\bibinfo{year}{2000}).

\bibitem[{\citenamefont{Rauscher et~al.}(2002)\citenamefont{Rauscher, Heger,
  Hoffman, and Woosley}}]{RHH02}
\bibinfo{author}{\bibfnamefont{T.}~\bibnamefont{Rauscher}},
  \bibinfo{author}{\bibfnamefont{A.}~\bibnamefont{Heger}},
  \bibinfo{author}{\bibfnamefont{R.}~\bibnamefont{Hoffman}}, \bibnamefont{and}
  \bibinfo{author}{\bibfnamefont{S.}~\bibnamefont{Woosley}},
  \bibinfo{journal}{Ap. J.} \textbf{\bibinfo{volume}{576}},
  \bibinfo{pages}{323} (\bibinfo{year}{2002}).

\bibitem[{\citenamefont{Heil et~al.}(2008{\natexlab{a}})\citenamefont{Heil,
  K{\"a}ppeler, Uberseder, Gallino, and Pignatari}}]{HKU08a}
\bibinfo{author}{\bibfnamefont{M.}~\bibnamefont{Heil}},
  \bibinfo{author}{\bibfnamefont{F.}~\bibnamefont{K{\"a}ppeler}},
  \bibinfo{author}{\bibfnamefont{E.}~\bibnamefont{Uberseder}},
  \bibinfo{author}{\bibfnamefont{R.}~\bibnamefont{Gallino}}, \bibnamefont{and}
  \bibinfo{author}{\bibfnamefont{M.}~\bibnamefont{Pignatari}},
  \bibinfo{journal}{Phys. Rev. C} \textbf{\bibinfo{volume}{77}},
  \bibinfo{pages}{015808} (\bibinfo{year}{2008}{\natexlab{a}}).

\bibitem[{\citenamefont{Heil et~al.}(2008{\natexlab{b}})\citenamefont{Heil,
  K{\"a}ppeler, Uberseder, Gallino, Bisterzo, and Pignatari}}]{HKU08b}
\bibinfo{author}{\bibfnamefont{M.}~\bibnamefont{Heil}},
  \bibinfo{author}{\bibfnamefont{F.}~\bibnamefont{K{\"a}ppeler}},
  \bibinfo{author}{\bibfnamefont{E.}~\bibnamefont{Uberseder}},
  \bibinfo{author}{\bibfnamefont{R.}~\bibnamefont{Gallino}},
  \bibinfo{author}{\bibfnamefont{S.}~\bibnamefont{Bisterzo}}, \bibnamefont{and}
  \bibinfo{author}{\bibfnamefont{M.}~\bibnamefont{Pignatari}},
  \bibinfo{journal}{Phys. Rev. C} \textbf{\bibinfo{volume}{78}},
  \bibinfo{pages}{025802} (\bibinfo{year}{2008}{\natexlab{b}}).

\bibitem[{\citenamefont{Pignatari et~al.}(2010)\citenamefont{Pignatari,
  Gallino, Heil, Wiescher, K{\"a}ppeler, Herwig, and Bisterzo}}]{PGH10}
\bibinfo{author}{\bibfnamefont{M.}~\bibnamefont{Pignatari}},
  \bibinfo{author}{\bibfnamefont{R.}~\bibnamefont{Gallino}},
  \bibinfo{author}{\bibfnamefont{M.}~\bibnamefont{Heil}},
  \bibinfo{author}{\bibfnamefont{M.}~\bibnamefont{Wiescher}},
  \bibinfo{author}{\bibfnamefont{F.}~\bibnamefont{K{\"a}ppeler}},
  \bibinfo{author}{\bibfnamefont{F.}~\bibnamefont{Herwig}}, \bibnamefont{and}
  \bibinfo{author}{\bibfnamefont{S.}~\bibnamefont{Bisterzo}},
  \bibinfo{journal}{Ap. J.} \textbf{\bibinfo{volume}{710}},
  \bibinfo{pages}{1557} (\bibinfo{year}{2010}).

\bibitem[{\citenamefont{Bao et~al.}(2000)\citenamefont{Bao, Beer, K{\"a}ppeler,
  Voss, Wisshak, and Rauscher}}]{BBK00}
\bibinfo{author}{\bibfnamefont{Z.}~\bibnamefont{Bao}},
  \bibinfo{author}{\bibfnamefont{H.}~\bibnamefont{Beer}},
  \bibinfo{author}{\bibfnamefont{F.}~\bibnamefont{K{\"a}ppeler}},
  \bibinfo{author}{\bibfnamefont{F.}~\bibnamefont{Voss}},
  \bibinfo{author}{\bibfnamefont{K.}~\bibnamefont{Wisshak}}, \bibnamefont{and}
  \bibinfo{author}{\bibfnamefont{T.}~\bibnamefont{Rauscher}},
  \bibinfo{journal}{Atomic Data Nucl. Data Tables}
  \textbf{\bibinfo{volume}{76}}, \bibinfo{pages}{70} (\bibinfo{year}{2000}).

\bibitem[{\citenamefont{Dillmann et~al.}(2009)\citenamefont{Dillmann, Plag,
  K{\"a}ppeler, and Rauscher}}]{DPK09}
\bibinfo{author}{\bibfnamefont{I.}~\bibnamefont{Dillmann}},
  \bibinfo{author}{\bibfnamefont{R.}~\bibnamefont{Plag}},
  \bibinfo{author}{\bibfnamefont{F.}~\bibnamefont{K{\"a}ppeler}},
  \bibnamefont{and} \bibinfo{author}{\bibfnamefont{T.}~\bibnamefont{Rauscher}},
  in \emph{\bibinfo{booktitle}{EFNUDAT Fast Neutrons - scientific workshop on
  neutron measurements, theory \& applications}}
  (\bibinfo{publisher}{JRC-IRMM}, \bibinfo{address}{Geel},
  \bibinfo{year}{2009}), \bibinfo{note}{http://www.kadonis.org}.

\bibitem[{\citenamefont{Bihain et~al.}(2004)\citenamefont{Bihain, Israelian,
  Rebolo, Bonifacio, and Molaro}}]{BIR04}
\bibinfo{author}{\bibfnamefont{G.}~\bibnamefont{Bihain}},
  \bibinfo{author}{\bibfnamefont{G.}~\bibnamefont{Israelian}},
  \bibinfo{author}{\bibfnamefont{R.}~\bibnamefont{Rebolo}},
  \bibinfo{author}{\bibfnamefont{P.}~\bibnamefont{Bonifacio}},
  \bibnamefont{and} \bibinfo{author}{\bibfnamefont{P.}~\bibnamefont{Molaro}},
  \bibinfo{journal}{Astron. Astrophys.} \textbf{\bibinfo{volume}{423}},
  \bibinfo{pages}{777786} (\bibinfo{year}{2004}).

\bibitem[{\citenamefont{Chen et~al.}(2004)\citenamefont{Chen, Nissen, and
  Zhao}}]{CNZ04}
\bibinfo{author}{\bibfnamefont{Y.}~\bibnamefont{Chen}},
  \bibinfo{author}{\bibfnamefont{P.}~\bibnamefont{Nissen}}, \bibnamefont{and}
  \bibinfo{author}{\bibfnamefont{G.}~\bibnamefont{Zhao}},
  \bibinfo{journal}{Astron. Astrophys.} \textbf{\bibinfo{volume}{425}},
  \bibinfo{pages}{697705} (\bibinfo{year}{2004}).

\bibitem[{\citenamefont{Nissen et~al.}(2004)\citenamefont{Nissen, Chen,
  Asplund, and Pettini}}]{NCA04}
\bibinfo{author}{\bibfnamefont{P.}~\bibnamefont{Nissen}},
  \bibinfo{author}{\bibfnamefont{Y.}~\bibnamefont{Chen}},
  \bibinfo{author}{\bibfnamefont{M.}~\bibnamefont{Asplund}}, \bibnamefont{and}
  \bibinfo{author}{\bibfnamefont{M.}~\bibnamefont{Pettini}},
  \bibinfo{journal}{Astron. Astrophys.} \textbf{\bibinfo{volume}{415}},
  \bibinfo{pages}{9931007} (\bibinfo{year}{2004}).

\bibitem[{\citenamefont{Beer and K{\"a}ppeler}(1980)}]{BeK80}
\bibinfo{author}{\bibfnamefont{H.}~\bibnamefont{Beer}} \bibnamefont{and}
  \bibinfo{author}{\bibfnamefont{F.}~\bibnamefont{K{\"a}ppeler}},
  \bibinfo{journal}{Phys. Rev. C} \textbf{\bibinfo{volume}{21}},
  \bibinfo{pages}{534} (\bibinfo{year}{1980}).

\bibitem[{\citenamefont{Patronis et~al.}(2004)\citenamefont{Patronis, Dababneh,
  Assimakopoulos, Gallino, Heil, K{\"a}ppeler, Karamanis, Koehler, Mengoni, and
  Plag}}]{PAK04}
\bibinfo{author}{\bibfnamefont{N.}~\bibnamefont{Patronis}},
  \bibinfo{author}{\bibfnamefont{S.}~\bibnamefont{Dababneh}},
  \bibinfo{author}{\bibfnamefont{P.}~\bibnamefont{Assimakopoulos}},
  \bibinfo{author}{\bibfnamefont{R.}~\bibnamefont{Gallino}},
  \bibinfo{author}{\bibfnamefont{M.}~\bibnamefont{Heil}},
  \bibinfo{author}{\bibfnamefont{F.}~\bibnamefont{K{\"a}ppeler}},
  \bibinfo{author}{\bibfnamefont{D.}~\bibnamefont{Karamanis}},
  \bibinfo{author}{\bibfnamefont{P.}~\bibnamefont{Koehler}},
  \bibinfo{author}{\bibfnamefont{A.}~\bibnamefont{Mengoni}}, \bibnamefont{and}
  \bibinfo{author}{\bibfnamefont{R.}~\bibnamefont{Plag}},
  \bibinfo{journal}{Phys. Rev. C} \textbf{\bibinfo{volume}{69}},
  \bibinfo{pages}{025803} (\bibinfo{year}{2004}).

\bibitem[{\citenamefont{O'Brien et~al.}(2003)\citenamefont{O'Brien, Dababneh,
  Heil, K{\"a}ppeler, Plag, Reifarth, Gallino, and Pignatari}}]{BDH03}
\bibinfo{author}{\bibfnamefont{S.}~\bibnamefont{O'Brien}},
  \bibinfo{author}{\bibfnamefont{S.}~\bibnamefont{Dababneh}},
  \bibinfo{author}{\bibfnamefont{M.}~\bibnamefont{Heil}},
  \bibinfo{author}{\bibfnamefont{F.}~\bibnamefont{K{\"a}ppeler}},
  \bibinfo{author}{\bibfnamefont{R.}~\bibnamefont{Plag}},
  \bibinfo{author}{\bibfnamefont{R.}~\bibnamefont{Reifarth}},
  \bibinfo{author}{\bibfnamefont{R.}~\bibnamefont{Gallino}}, \bibnamefont{and}
  \bibinfo{author}{\bibfnamefont{M.}~\bibnamefont{Pignatari}},
  \bibinfo{journal}{Phys. Rev. C} \textbf{\bibinfo{volume}{68}},
  \bibinfo{pages}{035801} (\bibinfo{year}{2003}).

\bibitem[{\citenamefont{Ratzel et~al.}(2004)\citenamefont{Ratzel, Arlandini,
  K{\"a}ppeler, Couture, Wiescher, Reifarth, Gallino, Mengoni, and
  Travaglio}}]{RAK04}
\bibinfo{author}{\bibfnamefont{U.}~\bibnamefont{Ratzel}},
  \bibinfo{author}{\bibfnamefont{C.}~\bibnamefont{Arlandini}},
  \bibinfo{author}{\bibfnamefont{F.}~\bibnamefont{K{\"a}ppeler}},
  \bibinfo{author}{\bibfnamefont{A.}~\bibnamefont{Couture}},
  \bibinfo{author}{\bibfnamefont{M.}~\bibnamefont{Wiescher}},
  \bibinfo{author}{\bibfnamefont{R.}~\bibnamefont{Reifarth}},
  \bibinfo{author}{\bibfnamefont{R.}~\bibnamefont{Gallino}},
  \bibinfo{author}{\bibfnamefont{A.}~\bibnamefont{Mengoni}}, \bibnamefont{and}
  \bibinfo{author}{\bibfnamefont{C.}~\bibnamefont{Travaglio}},
  \bibinfo{journal}{Phys. Rev. C} \textbf{\bibinfo{volume}{70}},
  \bibinfo{pages}{065803} (\bibinfo{year}{2004}).

\bibitem[{\citenamefont{Ratynski and K{\"a}ppeler}(1988)}]{RaK88}
\bibinfo{author}{\bibfnamefont{W.}~\bibnamefont{Ratynski}} \bibnamefont{and}
  \bibinfo{author}{\bibfnamefont{F.}~\bibnamefont{K{\"a}ppeler}},
  \bibinfo{journal}{Phys. Rev. C} \textbf{\bibinfo{volume}{37}},
  \bibinfo{pages}{595} (\bibinfo{year}{1988}).

\bibitem[{\citenamefont{de~Laeter et~al.}(2003)\citenamefont{de~Laeter,
  B{\"o}hlke, de~Bi\`evre, Hidaka, Peiser, Rosman, and Taylor}}]{DBD03}
\bibinfo{author}{\bibfnamefont{J.}~\bibnamefont{de~Laeter}},
  \bibinfo{author}{\bibfnamefont{J.}~\bibnamefont{B{\"o}hlke}},
  \bibinfo{author}{\bibfnamefont{P.}~\bibnamefont{de~Bi\`evre}},
  \bibinfo{author}{\bibfnamefont{H.}~\bibnamefont{Hidaka}},
  \bibinfo{author}{\bibfnamefont{H.}~\bibnamefont{Peiser}},
  \bibinfo{author}{\bibfnamefont{K.}~\bibnamefont{Rosman}}, \bibnamefont{and}
  \bibinfo{author}{\bibfnamefont{P.}~\bibnamefont{Taylor}},
  \bibinfo{journal}{Pure Appl. Chem.} \textbf{\bibinfo{volume}{75}},
  \bibinfo{pages}{683} (\bibinfo{year}{2003}).

\bibitem[{\citenamefont{Browne and Tuli}(2010)}]{BrT10}
\bibinfo{author}{\bibfnamefont{E.}~\bibnamefont{Browne}} \bibnamefont{and}
  \bibinfo{author}{\bibfnamefont{J.}~\bibnamefont{Tuli}},
  \bibinfo{journal}{Nucl. Data Sheets} \textbf{\bibinfo{volume}{111}},
  \bibinfo{pages}{2425 } (\bibinfo{year}{2010}).

\bibitem[{\citenamefont{Bhat and Tuli}(2000)}]{BhT00}
\bibinfo{author}{\bibfnamefont{M.}~\bibnamefont{Bhat}} \bibnamefont{and}
  \bibinfo{author}{\bibfnamefont{J.}~\bibnamefont{Tuli}},
  \bibinfo{journal}{Nucl. Data Sheets} \textbf{\bibinfo{volume}{90}},
  \bibinfo{pages}{269} (\bibinfo{year}{2000}).

\bibitem[{\citenamefont{Abusaleem and Singh}(2011)}]{AbS11}
\bibinfo{author}{\bibfnamefont{K.}~\bibnamefont{Abusaleem}} \bibnamefont{and}
  \bibinfo{author}{\bibfnamefont{B.}~\bibnamefont{Singh}},
  \bibinfo{journal}{Nuclear Data Sheets} \textbf{\bibinfo{volume}{112}},
  \bibinfo{pages}{133 } (\bibinfo{year}{2011}).

\bibitem[{\citenamefont{Huang}(2009)}]{Hua09}
\bibinfo{author}{\bibfnamefont{X.}~\bibnamefont{Huang}},
  \bibinfo{journal}{Nuclear Data Sheets} \textbf{\bibinfo{volume}{110}},
  \bibinfo{pages}{2533 } (\bibinfo{year}{2009}).

\bibitem[{\citenamefont{Dababneh et~al.}(2004)\citenamefont{Dababneh, Patronis,
  Assimakopoulos, G{\"o}rres, Heil, K{\"a}ppeler, Karamanis, O'Brien, and
  Reifarth}}]{DPA04}
\bibinfo{author}{\bibfnamefont{S.}~\bibnamefont{Dababneh}},
  \bibinfo{author}{\bibfnamefont{N.}~\bibnamefont{Patronis}},
  \bibinfo{author}{\bibfnamefont{P.}~\bibnamefont{Assimakopoulos}},
  \bibinfo{author}{\bibfnamefont{J.}~\bibnamefont{G{\"o}rres}},
  \bibinfo{author}{\bibfnamefont{M.}~\bibnamefont{Heil}},
  \bibinfo{author}{\bibfnamefont{F.}~\bibnamefont{K{\"a}ppeler}},
  \bibinfo{author}{\bibfnamefont{D.}~\bibnamefont{Karamanis}},
  \bibinfo{author}{\bibfnamefont{S.}~\bibnamefont{O'Brien}}, \bibnamefont{and}
  \bibinfo{author}{\bibfnamefont{R.}~\bibnamefont{Reifarth}},
  \bibinfo{journal}{Nucl. Instr. Meth.} \textbf{\bibinfo{volume}{A517}},
  \bibinfo{pages}{230 } (\bibinfo{year}{2004}).

\bibitem[{\citenamefont{Levkovskii}(1958)}]{Lev58}
\bibinfo{author}{\bibfnamefont{V.}~\bibnamefont{Levkovskii}},
  \bibinfo{journal}{Atomnaya Energ.} \textbf{\bibinfo{volume}{4}},
  \bibinfo{pages}{79} (\bibinfo{year}{1958}).

\bibitem[{\citenamefont{Thwaites and Pratt}(1961)}]{ThP61}
\bibinfo{author}{\bibfnamefont{T.}~\bibnamefont{Thwaites}} \bibnamefont{and}
  \bibinfo{author}{\bibfnamefont{W.}~\bibnamefont{Pratt}},
  \bibinfo{journal}{Phys. Rev.} \textbf{\bibinfo{volume}{124}},
  \bibinfo{pages}{1526} (\bibinfo{year}{1961}).

\bibitem[{\citenamefont{Sonnino et~al.}(1964)\citenamefont{Sonnino, Eichler,
  and Amiel}}]{SEA64}
\bibinfo{author}{\bibfnamefont{T.}~\bibnamefont{Sonnino}},
  \bibinfo{author}{\bibfnamefont{E.}~\bibnamefont{Eichler}}, \bibnamefont{and}
  \bibinfo{author}{\bibfnamefont{S.}~\bibnamefont{Amiel}},
  \bibinfo{journal}{Nucl. Phys. A} \textbf{\bibinfo{volume}{54}},
  \bibinfo{pages}{568 } (\bibinfo{year}{1964}).

\bibitem[{\citenamefont{Reifarth et~al.}(2003)\citenamefont{Reifarth,
  Arlandini, Heil, K{\"a}ppeler, Sedyshev, Herman, Rauscher, Gallino, and
  Travaglio}}]{RAH03}
\bibinfo{author}{\bibfnamefont{R.}~\bibnamefont{Reifarth}},
  \bibinfo{author}{\bibfnamefont{C.}~\bibnamefont{Arlandini}},
  \bibinfo{author}{\bibfnamefont{M.}~\bibnamefont{Heil}},
  \bibinfo{author}{\bibfnamefont{F.}~\bibnamefont{K{\"a}ppeler}},
  \bibinfo{author}{\bibfnamefont{P.}~\bibnamefont{Sedyshev}},
  \bibinfo{author}{\bibfnamefont{M.}~\bibnamefont{Herman}},
  \bibinfo{author}{\bibfnamefont{T.}~\bibnamefont{Rauscher}},
  \bibinfo{author}{\bibfnamefont{R.}~\bibnamefont{Gallino}}, \bibnamefont{and}
  \bibinfo{author}{\bibfnamefont{C.}~\bibnamefont{Travaglio}},
  \bibinfo{journal}{Ap. J.} \textbf{\bibinfo{volume}{582}},
  \bibinfo{pages}{1251 } (\bibinfo{year}{2003}).

\bibitem[{\citenamefont{Reifarth et~al.}(2009)\citenamefont{Reifarth, Heil,
  K{\"a}ppeler, and Plag}}]{RHK09}
\bibinfo{author}{\bibfnamefont{R.}~\bibnamefont{Reifarth}},
  \bibinfo{author}{\bibfnamefont{M.}~\bibnamefont{Heil}},
  \bibinfo{author}{\bibfnamefont{F.}~\bibnamefont{K{\"a}ppeler}},
  \bibnamefont{and} \bibinfo{author}{\bibfnamefont{R.}~\bibnamefont{Plag}},
  \bibinfo{journal}{Nucl. Instr. Meth. A} \textbf{\bibinfo{volume}{608}},
  \bibinfo{pages}{139} (\bibinfo{year}{2009}).

\bibitem[{\citenamefont{Macklin}(1982)}]{Mac82a}
\bibinfo{author}{\bibfnamefont{R.}~\bibnamefont{Macklin}}
  (\bibinfo{year}{1982}), \bibinfo{note}{private communication to Mughabghab,
  S.F. (1982), see also www.nndc.bnl.gov/nndc/ EXFOR 12720.002}.

\bibitem[{\citenamefont{Dillmann
  et~al.}(2006{\natexlab{a}})\citenamefont{Dillmann, Heil, K{\"a}ppeler, Plag,
  Rauscher, and Thielemann}}]{DHK06}
\bibinfo{author}{\bibfnamefont{I.}~\bibnamefont{Dillmann}},
  \bibinfo{author}{\bibfnamefont{M.}~\bibnamefont{Heil}},
  \bibinfo{author}{\bibfnamefont{F.}~\bibnamefont{K{\"a}ppeler}},
  \bibinfo{author}{\bibfnamefont{R.}~\bibnamefont{Plag}},
  \bibinfo{author}{\bibfnamefont{T.}~\bibnamefont{Rauscher}}, \bibnamefont{and}
  \bibinfo{author}{\bibfnamefont{F.-K.} \bibnamefont{Thielemann}}, in
  \emph{\bibinfo{booktitle}{Capture Gamma-Ray Spectroscopy and Related
  Topics}}, edited by \bibinfo{editor}{\bibfnamefont{A.}~\bibnamefont{Woehr}}
  \bibnamefont{and}
  \bibinfo{editor}{\bibfnamefont{A.}~\bibnamefont{Aprahamian}}
  (\bibinfo{publisher}{AIP}, \bibinfo{address}{New York},
  \bibinfo{year}{2006}{\natexlab{a}}), AIP Conference Series 819, p.
  \bibinfo{pages}{123}, \bibinfo{note}{http://www.kadonis.org}.

\bibitem[{\citenamefont{Garg et~al.}(1981)\citenamefont{Garg, Tikku, Halperin,
  and Macklin}}]{GTH81}
\bibinfo{author}{\bibfnamefont{J.}~\bibnamefont{Garg}},
  \bibinfo{author}{\bibfnamefont{V.}~\bibnamefont{Tikku}},
  \bibinfo{author}{\bibfnamefont{J.}~\bibnamefont{Halperin}}, \bibnamefont{and}
  \bibinfo{author}{\bibfnamefont{R.}~\bibnamefont{Macklin}},
  \bibinfo{journal}{Phys. Rev. C} \textbf{\bibinfo{volume}{23}},
  \bibinfo{pages}{683} (\bibinfo{year}{1981}).

\bibitem[{\citenamefont{Garg et~al.}(1982)\citenamefont{Garg, Tikku, , Harvey,
  Halperin, and Macklin}}]{GTH82}
\bibinfo{author}{\bibfnamefont{J.}~\bibnamefont{Garg}},
  \bibinfo{author}{\bibfnamefont{V.}~\bibnamefont{Tikku}}, ,
  \bibinfo{author}{\bibfnamefont{J.}~\bibnamefont{Harvey}},
  \bibinfo{author}{\bibfnamefont{J.}~\bibnamefont{Halperin}}, \bibnamefont{and}
  \bibinfo{author}{\bibfnamefont{R.}~\bibnamefont{Macklin}},
  \bibinfo{journal}{Phys. Rev. C} \textbf{\bibinfo{volume}{25}},
  \bibinfo{pages}{1808} (\bibinfo{year}{1982}).

\bibitem[{\citenamefont{Murty et~al.}(1973)\citenamefont{Murty, Siddappa, and
  Rao}}]{MSR73}
\bibinfo{author}{\bibfnamefont{M.}~\bibnamefont{Murty}},
  \bibinfo{author}{\bibfnamefont{K.}~\bibnamefont{Siddappa}}, \bibnamefont{and}
  \bibinfo{author}{\bibfnamefont{J.}~\bibnamefont{Rao}}, \bibinfo{journal}{J.
  Phys. Soc. Jap.} \textbf{\bibinfo{volume}{35}}, \bibinfo{pages}{8}
  (\bibinfo{year}{1973}).

\bibitem[{\citenamefont{Rauscher and Thielemann}(2000)}]{RaT00}
\bibinfo{author}{\bibfnamefont{T.}~\bibnamefont{Rauscher}} \bibnamefont{and}
  \bibinfo{author}{\bibfnamefont{F.-K.} \bibnamefont{Thielemann}},
  \bibinfo{journal}{Atomic Data Nucl. Data Tables}
  \textbf{\bibinfo{volume}{75}}, \bibinfo{pages}{1} (\bibinfo{year}{2000}).

\bibitem[{\citenamefont{Dillmann
  et~al.}(2006{\natexlab{b}})\citenamefont{Dillmann, Plag, K{\"a}ppeler, and
  Rauscher}}]{DPK06}
\bibinfo{author}{\bibfnamefont{I.}~\bibnamefont{Dillmann}},
  \bibinfo{author}{\bibfnamefont{R.}~\bibnamefont{Plag}},
  \bibinfo{author}{\bibfnamefont{F.}~\bibnamefont{K{\"a}ppeler}},
  \bibnamefont{and} \bibinfo{author}{\bibfnamefont{T.}~\bibnamefont{Rauscher}},
  in \emph{\bibinfo{booktitle}{International Symposium on Nuclear Astrophysics,
  Nuclei in the Cosmos - IX}}, edited by
  \bibinfo{editor}{\bibfnamefont{A.~e.~a.} \bibnamefont{Mengoni}}
  (\bibinfo{publisher}{SISSA}, \bibinfo{address}{Trieste},
  \bibinfo{year}{2006}{\natexlab{b}}), PoS - Proceedings of Science, ISSN
  18248039, p. \bibinfo{pages}{article 090},
  \bibinfo{note}{http://pos.sissa.it}.

\bibitem[{\citenamefont{Harris}(1981)}]{Har81}
\bibinfo{author}{\bibfnamefont{M.}~\bibnamefont{Harris}}, \bibinfo{journal}{Ap.
  Space Sci.} \textbf{\bibinfo{volume}{77}}, \bibinfo{pages}{357}
  (\bibinfo{year}{1981}).

\bibitem[{\citenamefont{Woosley et~al.}(1978)\citenamefont{Woosley, Fowler,
  Holmes, and Zimmerman}}]{WFH78}
\bibinfo{author}{\bibfnamefont{S.}~\bibnamefont{Woosley}},
  \bibinfo{author}{\bibfnamefont{W.}~\bibnamefont{Fowler}},
  \bibinfo{author}{\bibfnamefont{J.}~\bibnamefont{Holmes}}, \bibnamefont{and}
  \bibinfo{author}{\bibfnamefont{B.}~\bibnamefont{Zimmerman}},
  \bibinfo{journal}{Atomic Data Nucl. Data Tables}
  \textbf{\bibinfo{volume}{22}}, \bibinfo{pages}{371} (\bibinfo{year}{1978}).

\bibitem[{\citenamefont{Goriely}(2002)}]{Gor02}
\bibinfo{author}{\bibfnamefont{S.}~\bibnamefont{Goriely}}, \bibinfo{type}{Tech.
  Rep.} (\bibinfo{year}{2002}), \bibinfo{note}{http://www-astro.ulb.ac.be}.

\bibitem[{\citenamefont{Goriely}(2005)}]{Gor05}
\bibinfo{author}{\bibfnamefont{S.}~\bibnamefont{Goriely}}, \bibinfo{type}{Tech.
  Rep.} (\bibinfo{year}{2005}), \bibinfo{note}{http://www-astro.ulb.ac.be}.

\bibitem[{\citenamefont{Jaag}(1991)}]{Jaa91}
\bibinfo{author}{\bibfnamefont{S.}~\bibnamefont{Jaag}}, \bibinfo{type}{Tech.
  Rep.}, \bibinfo{institution}{Forschungszentrum Karlsruhe}
  (\bibinfo{year}{1991}).

\bibitem[{\citenamefont{Raiteri
  et~al.}(1991{\natexlab{b}})\citenamefont{Raiteri, Busso, Gallino, Picchio,
  and Pulone}}]{RBG91a}
\bibinfo{author}{\bibfnamefont{C.}~\bibnamefont{Raiteri}},
  \bibinfo{author}{\bibfnamefont{M.}~\bibnamefont{Busso}},
  \bibinfo{author}{\bibfnamefont{R.}~\bibnamefont{Gallino}},
  \bibinfo{author}{\bibfnamefont{G.}~\bibnamefont{Picchio}}, \bibnamefont{and}
  \bibinfo{author}{\bibfnamefont{L.}~\bibnamefont{Pulone}},
  \bibinfo{journal}{Ap. J.} \textbf{\bibinfo{volume}{367}},
  \bibinfo{pages}{228} (\bibinfo{year}{1991}{\natexlab{b}}).

\end{thebibliography}
\end{document}